\documentclass{PoS}

\usepackage{subcaption}

\usepackage{amsmath, amssymb, mathrsfs}

\usepackage[utf8]{inputenc}

\newcommand{\mevcc}{MeV/c$^2$}
\newcommand{\gevcc}{GeV/c$^2$}
\newcommand{\posi}{e^{+}}
\newcommand{\ele}{e^{-}}
\newcommand{\epem}{\posi\ele}
\newcommand{\emem}{\ele\ele}

\newcommand{\aprime}{A^\prime}

\title{Search for a Dark Photon in Electro-Produced $\epem$ Pairs with the 
       Heavy Photon Search Experiment at JLab}

\author{Omar Moreno$^{ab}$\footnote{Speaker} \hspace{0.05cm} and \hspace{0.05 cm}
        Matthew Solt,$^a$\footnotemark[1] \hspace{0.05cm} for the HPS Collaboration \\
        \llap{$^a$} SLAC National Accelerator Laboratory, 
                    Stanford  University,  Stanford,  CA  94309,  USA \\
        \llap{$^b$} Santa  Cruz  Institute  for  Particle  Physics,
                    University  of  California,  Santa  Cruz,  CA  95064,  USA \\
        E-mail: \email{omoreno@slac.stanford.edu}, \email{mrsolt@slac.stanford.edu}
        }

\abstract{The Heavy Photon Search experiment took its first data in a 2015
          engineering run using a 1.056 GeV, 50 nA electron beam provided by
          CEBAF at the Thomas Jefferson National Accelerator Facility,
          searching for an electro-produced dark photon. Using 1.7 days 
          (1170 nb$^{-1}$) of data, a search for a resonance in the $\epem$ 
          invariant mass distribution between 19 and 81~\mevcc~
          showed no evidence of dark photon decays above the large QED
          background, confirming earlier searches and demonstrating the full 
          functionality of the experiment. Upper limits on the square of the
          coupling of the dark photon to the Standard Model photon are set at the
          level of 6$\times$10$^{-6}$. In addition, a search for displaced dark 
          photon decays did not rule out any territory but resulted in a reliable
          analysis procedure that will probe hitherto unexplored parameter space with
          future, higher luminosity runs.
        }

\FullConference{XXXIX International Conference on High Energy Physics \\
		        4-11 July 2018\\
		        Seoul, South Korea}

\ShortTitle{Search for a Dark Photon with HPS}
		        
\begin{document}

\section{Introduction}\label{sec:intro}

The existence of dark matter (DM) has been firmly established through 
its gravitational interaction as well as measurements of the power
spectrum of the Cosmic Microwave Background.  For a comprehensive review, 
see~\cite{Bertone:2004pz}. Even with such overwhelming evidence, the exact 
particle nature of DM continues to elude us. One of the simplest possibilities
envisions DM as originating as a thermal relic from the hot early Universe.
In such a scenario, the correct thermal relic abundance observed today can
be achieved only if DM has a small non-gravitational interaction with 
the Standard Model (SM) whose interaction rate exceeds the Hubble expansion
rate at some point in the early Universe.  Such a mechanism for generating
the thermal DM abundance not only leads to a minimum annihilation rate (
$\langle \sigma v \rangle \sim 10^{-26}$~cm$^3$s$^{-1}$) which is required to 
avoid an overabundance, but also defines a clear mass range 
($\sim$MeV - 10 TeV) for such candidates. This sets a region of phase space that
has to be experimentally probed in order to rule out thermal DM. 

Traditionally, searches for thermal DM have focused on 
Weakly Interacting Massive Particles (WIMPs) with masses between $\sim$GeV and 10 TeV, 
motivated by the connection between WIMPs and supersymmetry (SUSY) which also
predicts a DM candidate with similar properties. However, decades of direct 
and indirect searches for WIMPs have ruled out large regions of WIMP parameter space, and
next generation experiments (e.g. SuperCDMS, LZ) will probe large additional regions 
of parameter space.

Sub-GeV or light DM (LDM), in the broad vicinity of the weak scale, is a 
natural and simple generalization of WIMPs that has been difficult to test 
using experiments designed to probe WIMPs.  However, LDM requires the existence 
of a new force in order to achieve the correct thermal relic 
abundance~\cite{Lee:1977ua}. Such a candidate is well-motivated by 
``hidden sector'' scenarios which envision DM as having its own forces and
interactions which do not interact directly with 
SM~\cite{Alexander:2016aln, Battaglieri:2017aum}.  In the simplest hidden sector
models, DM is charged under a new $U(1)'$ gauge field mediated by a $U(1)'$ gauge
boson, $\aprime$ (``heavy'' or ``dark photon''). A heavy photon kinetically mixes
with a $U(1)_{Y}$ (hypercharge) gauge boson with a strength $\epsilon$, in turn 
inducing an effective coupling to electric charge.

Because of this effective coupling, heavy photons 
can be produced in a process analogous to bremsstrahlung radiation, subsequently 
decaying to charged lepton pairs. The Heavy Photon Search (HPS) experiment is a
fixed target experiment that utilizes this mechanism to produce heavy photons
using an intense electron beam incident on a thin tungsten target and then
detecting the $\epem$ decay products.  Specifically, the HPS experiment was 
designed to make use of such a production mechanism to search for a heavy photon
using two methods. The first is a resonance search for an excess above a large
QED background.  Such a search is only sensitive to large kinetic mixing strengths. 
For a sufficiently small kinetic mixing strengths, a second analysis can be 
performed by searching for a displaced vertex signature 
(in the range $\sim$1 - 10 cm) past a background of prompt QED tridents.
Using these search techniques,  HPS can explore the $\aprime$ mass range of 
19 ~\mevcc ~to 500 ~\mevcc. 

During a short engineering run in the spring of 2015, the HPS detector was installed,
commissioned, and operated in the experimental Hall B alcove at Jefferson Laboratory using a 
1.056 GeV, 50 nA beam provided by the Continuous Electron Beam Accelerator 
Facility (CEBAF). These proceedings report both the resonance and displaced vertex
searches for the 2015 engineering run in which a total of 1170 nb$^{-1}$ 
was collected. A more detailed description 
of the resonance search can be found in ~\cite{Adrian:2018scb}.

\section{Detector Overview}\label{sec:detector}

At the energies at which the HPS experiment is operating, 
the electro-produced $\aprime$ will carry most of the incident 
beam energy.  Consequently, the $\epem$ decay products are highly
boosted and necessitate a detector with a very forward acceptance
that can be placed in close proximity to the target.  Maximizing
the acceptance requires placing the detector close to the beam
plane, encroaching on a ``dead zone'' which is occupied by 
an intense flux of multiple Coulomb scattered beam particles along
with radiative secondaries originating from the target. In
order to avoid additional background from beam-gas interactions, 
the detector needs to be operated in vacuum.  Finally, minimizing
the material budget of the active area of the detector is essential
in reducing the multiple scattering that dominates both the 
mass and vertex resolutions that determine the experimental
sensitivity. 

\begin{figure}[th]
            \centering
            \includegraphics[width=.6\linewidth]{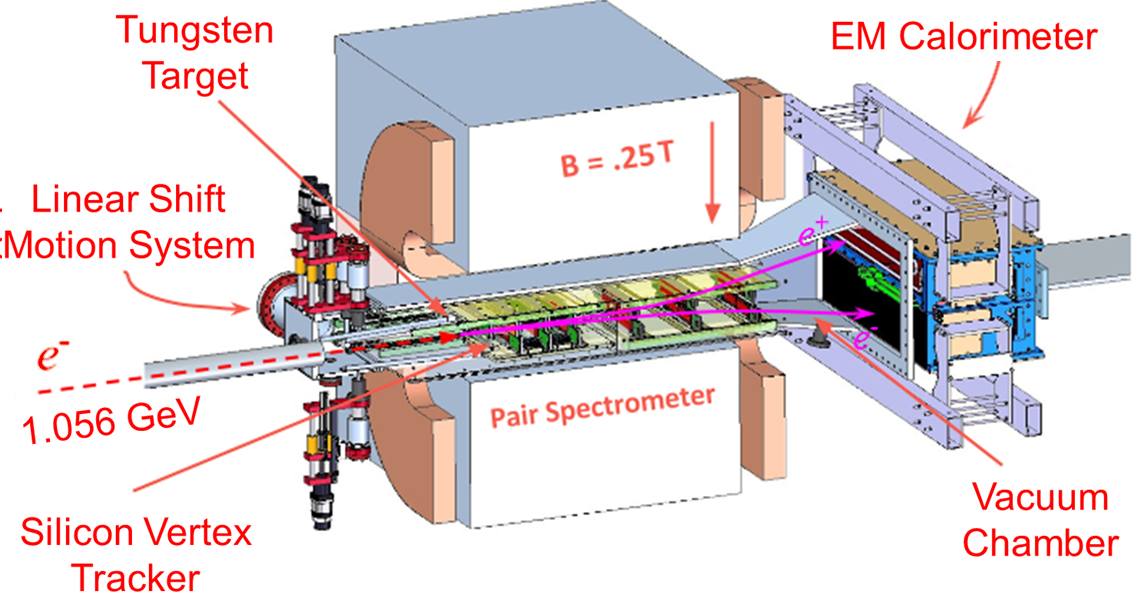}
            \caption{
                Schematic of the Heavy Photon Search Detector used during the 2015 engineering run.}
            \label{fig:detector}
        \end{figure}

With these design principles in mind, HPS utilizes a compact, large 
acceptance forward spectrometer consisting of a silicon vertex 
tracker (SVT) along with a lead tungstate electromagnetic calorimeter
(ECal) read out with high rate front end electronics.  The SVT is installed
inside a vacuum chamber immediately downstream of a thin (0.125\%) 
tungsten target. The vacuum chamber resides within an analyzing magnet 
providing a 0.24 Tesla field perpendicular to the beam plane, allowing for 
the precise measurement of track momenta. The ECal, placed downstream 
of the tracker, provides the primary trigger for the experiment and is 
also used for electron identification.  Together, both subsystems provide
the complete kinematic information required to reconstruct heavy 
photons. An overview of the HPS detector is shown in Figure~\ref{fig:detector}.

The SVT consists of two halves of six measurement stations enchroaching the beam plane. 
Each station consists of two silicon sensors where one is oriented at a small stereo
angle (50 or 100 mrad) in order to enable a full 3D hit reconstruction. 
The first silicon layer is located 0.5 mm from the beam plane, for the best possible
forward acceptance at 15 mrad, and is positioned about 10 cm downstream for the best
possible vertex resolution.
The silicon sensors are thin ($\sim$300 $\mu$m) in order to reduce multiple scattering.

During the 2015 engineering run, the HPS detector performed as expected with
momentum resolution, mass resolution, and vertex resolutions all in agreement with
predicted values.  For a detailed overview of the performance of the detector 
subsystems, see ~\cite{Balossino:2016nly, Baltzell:2016eee}.
\section{Resonance Search} \label{sec:bumphunt}

Searching for a heavy photon resonance requires accurate 
reconstruction of the $\epem$ invariant mass spectrum; rejection
of background events due to converted wide angle bremsstrahlung (WAB)
events, non-radiative
tridents from the Bethe-Heitler process, and occasional 
accidental $\epem$ pairs; and efficient selection of $\aprime$ 
candidates. Selecting $\aprime$ candidates is equivalent to selecting
radiative tridents since they have identical kinematics for a given 
mass. In order to perform a blind search, the event selection  was 
optimized using $\sim$10\% of the 2015 engineering run dataset.

Heavy photon candidates are created from pairs of electron and positron
tracks, one in each half of the SVT, each of which point to an energy  
cluster in the ECal.  Each track must pass loose quality requirements
and have a reconstructed momentum less than 75\% of the beam energy
(0.788 \gevcc) to reject scattered beam electrons.  The background from 
accidental pairs was reduced to less than 1\% by requiring the time between 
the ECal clusters be less than 2 ns and the time between a track and the
corresponding cluster be less than 5.8 ns.
        
Trident production, comprised of radiative and Bethe-Heitler processes, is
the main physics background to HPS. Although the Bethe-Heitler
diagram dominates over the radiative process, its different kinematics 
can be used to reduce its contribution to the final event sample. 
Specifically, at higher pair energies, the Bethe-Heitler process is 
not enhanced. 
As a result, the contribution of the Bethe-Heitler
process to the final event sample can be minimized by requiring the
momentum sum of the $\epem$ pair to be greater than 80\% of the beam
energy (0.84 \gevcc), where the radiative tridents are peaked.
        
The other significant source of background arises from converted WAB 
events in which the bremsstrahlung photon is emitted at a large angle
($>15$ mrad), converts in the target or first two layers of the SVT, and 
gives rise to a detected positron in the opposite half of the detector 
from the recoiling incoming electron.  The converted WAB background
was substantially reduced by requiring that both tracks have 
hits in the first two layers of the SVT.  Furthermore, applying 
additional requirements on the transverse momentum asymmetry between 
the $\epem$ ($<$ .47) and the transverse distance of closest approach
to the beam spot of the positron track ($<$ 1.1 mm) reduces the 
contamination from converted WABs down to 12\%. 

A prompt heavy photon is expected to appear as a Gaussian-shaped resonance
above the $\epem$ invariant mass spectrum, centered on the $\aprime$ mass
and with a width, $\sigma_{m_{\aprime}}$, characterized by the 
experimental mass resolution.  Calibration of the $\aprime$ mass scale
and resolution is done by using M\o ller scattering ($\emem \rightarrow \emem$)
events. The resulting parametrization is used as an input to the resonance search. 

Since the mass of the putative $\aprime$ is unknown a priori, the entire
$\epem$ invariant mass spectrum is scanned for any significant peaks.
This search is performed in a broad mass window around each candidate
mass, repeated in 0.5 MeV steps between 19 and 81 MeV. Within the window, 
the distribution of events is modeled using the 
probability distribution function
\begin{equation}
    P(m_{\epem}) = 
        \mu \cdot \phi(m_{\epem} | m_{\aprime}, \sigma_{m_{\aprime}}) 
            + B\cdot \exp(p(m_{\epem} | \mathbf{t}))
\end{equation}
where $m_{\epem}$ is the $\epem$ invariant mass, $\mu$ is the signal yield, 
$B$ is the number of background events within the window,  
$\phi(m_{\epem} | m_{\aprime}, \sigma_{m_{\aprime}})$ is a Gaussian 
probability distribution describing the signal and 
$p(m_{\epem} | \mathbf{t})$ is a Chebyshev polynomial of the first
kind with coefficients $\mathbf{t} = (t_{1}, ... t_{j})$ that is used 
to describe the background shape.  From optimization studies, a 
5th (3rd) order Chebyshev polynomial was found to best describe 
the background below (above) 39 MeV.  Note that $m_{\aprime}$ and
$\sigma_{m_{\aprime}}$ are set to the $\aprime$ mass hypothesis 
and experimental mass resolution, respectively.  Estimating the 
signal yield, the background normalization, and the background 
shape parameters within a window is done with a binned maximum 
likelihood fit using a bin width of 0.05 MeV.  A detailed 
discussion of the procedures followed can be found in \cite{Cowan:2010js}.
 Briefly, the log of the ratio of likelihoods for the 
 background-only fit to that of the best signal-plus-background fit 
 provides a test statistic from which the $p$-value can be calculated,
 giving the probability that the observed signal is a statistical 
 fluctuation.  The $p$-value is corrected for the 
 ``Look Elsewhere Effect'' (LEE) by performing simulated resonance 
searches on 4,000 pseudo data sets~\cite{Gross:2010qma}.

\begin{figure}[h]
    \centering
    \includegraphics[width=.49\linewidth]{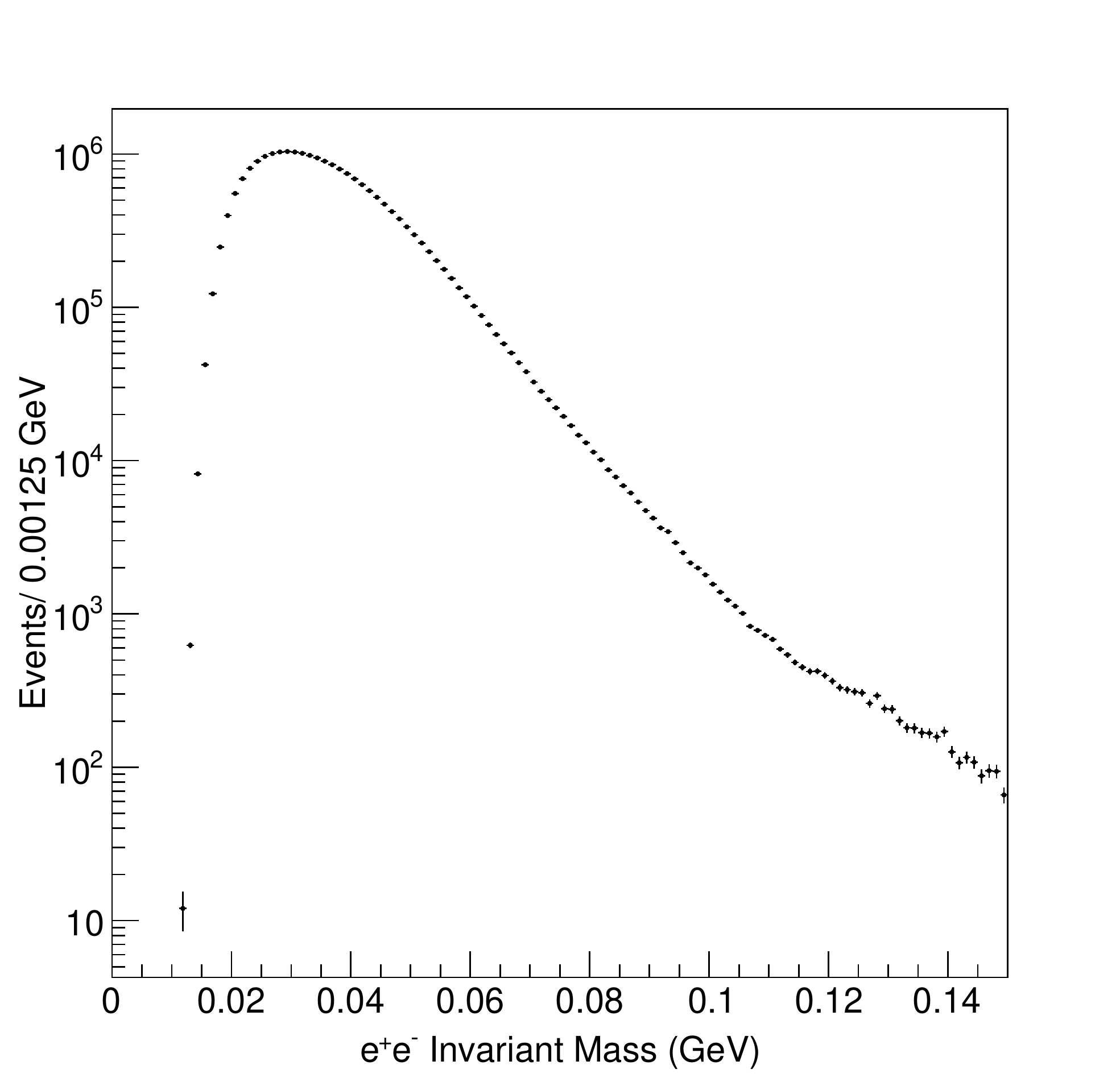}
    \includegraphics[width=.49\linewidth]{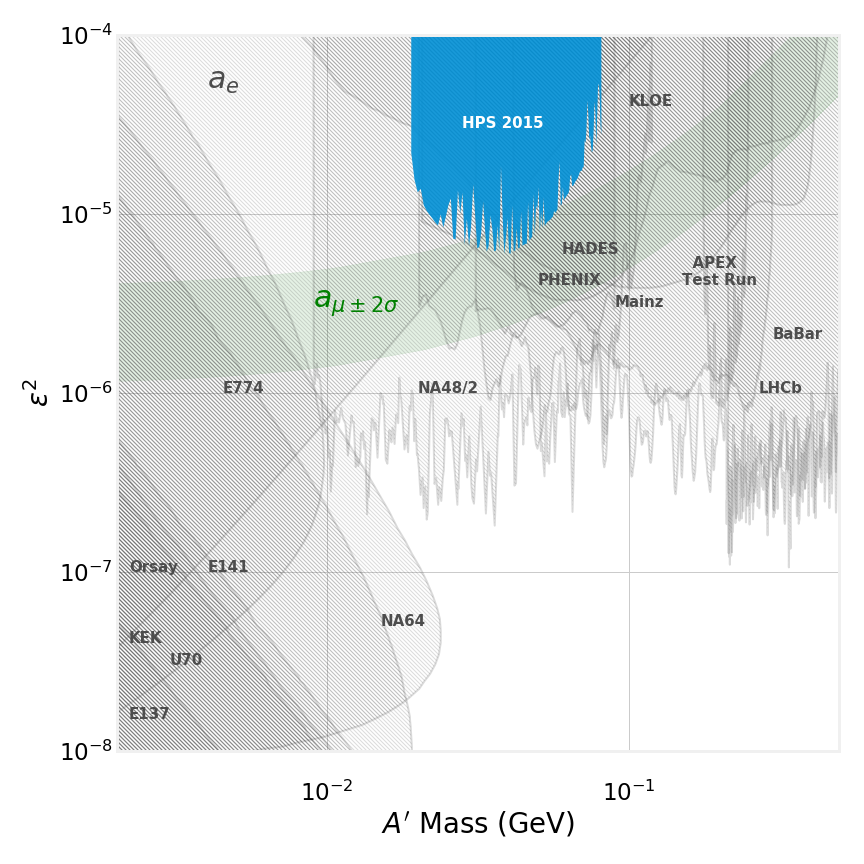}
    \caption{
        On the left, the distribution of $\epem$ invariant masses, events per
        1.25 MeV mass bin vs. mass.  On the right, 
        the 95\% C.L. power-constrained~\cite{Cowan:2011an} upper limits on 
        $\epsilon^2$ versus $\aprime$ mass obtained in this analysis. A limit
        at the level of 6$\times$ 10$^{-6}$ is set. Existing limits from beam
        dump, collider and fixed target experiments are also shown.  
        The region labeled ``$a_e$'' is an exclusion based on the electron 
        $g-2$.
        The green band 
        labeled ``$a_{\mu} \pm 2\sigma$'' represents the region that an $A'$ 
        can be used to explain the discrepancy between the measured and 
        calculated muon anomalous magnetic moment.
        A comprehensive review of all exclusions can be found in~\cite{Alexander:2016aln}. 
    }          
    \label{fig:epsilon_upper_limit}
\end{figure}

A search for a resonance in the $\epem$ invariant mass spectrum,
shown in Figure~\ref{fig:epsilon_upper_limit}, found no evidence of an $\aprime$ signal.
The most significant signal was observed at 37.7 MeV and has a 
local $p$-value of 0.17\%.  After accounting for the LEE correction, 
the most significant $p$-value is found to have a global 
$p$-value of 17\% corresponding to less than 2$\sigma$ in significance.
Since no significant signals were found, a  95\% C.L. upper limit is set, 
power-constrained~\cite{Cowan:2011an} to the expected limit. 
        
The proportionality between $\aprime$ and radiative trident 
production allows the normalization of the $\aprime$ rate to the 
measured rate of trident production~\cite{Bjorken:2009mm}.  This leads
to a relation that allows the signal upper limit, $S_{\text{up}}$, to be 
related to the $\aprime$ coupling strength as 
\begin{equation} \label{eps_coupling}
    \epsilon^2 = \left (\frac{S_{\text{up}}/m_{A'}}{
                f\Delta B/\Delta m} \right) 
            \left(\frac{2 N_{eff} \alpha}{3 \pi} \right)
\end{equation}
where $N_{eff}$ is the number of decay channels kinematically accessible
(=1 for HPS searches below the dimuon threshold), $\Delta B/\Delta m$
is the number of background events per MeV, $\alpha$ is the fine structure
constant and $f$ = 8.5\% is the 
fraction of radiative trident events comprising the background.  Using
equation \ref{eps_coupling}, the limits on $\epsilon$ set by HPS are
shown on Figure \ref{fig:epsilon_upper_limit}. 
        
The reach shown in Figure \ref{fig:epsilon_upper_limit} includes all 
statistical and  systematic uncertainties. The main systematic 
uncertainties on the signal yields arise from the uncertainty in the mass
resolution (3\%) and biases observed in the fit due to the background 
and signal parameterization (1.3-1.5\%, depending on mass). When scaling
the extracted signal yield upper limits to a limit on $\epsilon$, the
primary systematic uncertainty in the radiative fraction is due to 
the unknown composition of the final $\epem$ sample (7\%).  Many
other possible sources of systematic uncertainty were investigated and
accounted for but contribute negligibly to the result. 
\section{Displaced Vertex Search}\label{sec:vertex}

Searching for a long-lived heavy photon requires precise reconstruction of the vertex position of the $e^+e^-$ pair, rejection of background events that reconstruct significantly downstream of the target, and sufficient signal efficiency for a displaced $\aprime$. The precision of the vertex position is limited by multiple scattering in the SVT, particularly the first layer. The event selection is optimized for choosing well-reconstructed events (good tracks and vertex reconstruction) and tries to eliminate events which arise from scatters that could produce downstream reconstructed vertices. From $A^{\prime}$ simulation, cuts that improve our selection of events that have favorable kinematics were implemented for selecting heavy photons. 

Much of the initial event selection is similar to the resonance search in Section~\ref{sec:bumphunt}, such as timing cuts on tracks and ECal clusters, track quality requirements, $e^+e^-$ pairs in opposite halves of the SVT, tracks matched to ECal clusters, rejection of scattered beam electrons and WABs, and the radiative cut to select $e^+e^-$ pairs with energy sum near the beam energy. 
Specifically for the vertexing analysis, excellent vertex quality (based on the DOCA of the $e^+$ and $e^-$ tracks) with total momentum that points back to the beam spot is required. In order to reduce the effects of layer 1 mis-hits which could reconstruct a false vertex downstream of the target, an isolation cut that compares the distance between first layer hits and the impact parameter at the target was implemented, and tracks that share 5 hits with another track were removed.

\begin{figure}[th]
            \centering
            \includegraphics[width=.45\linewidth]{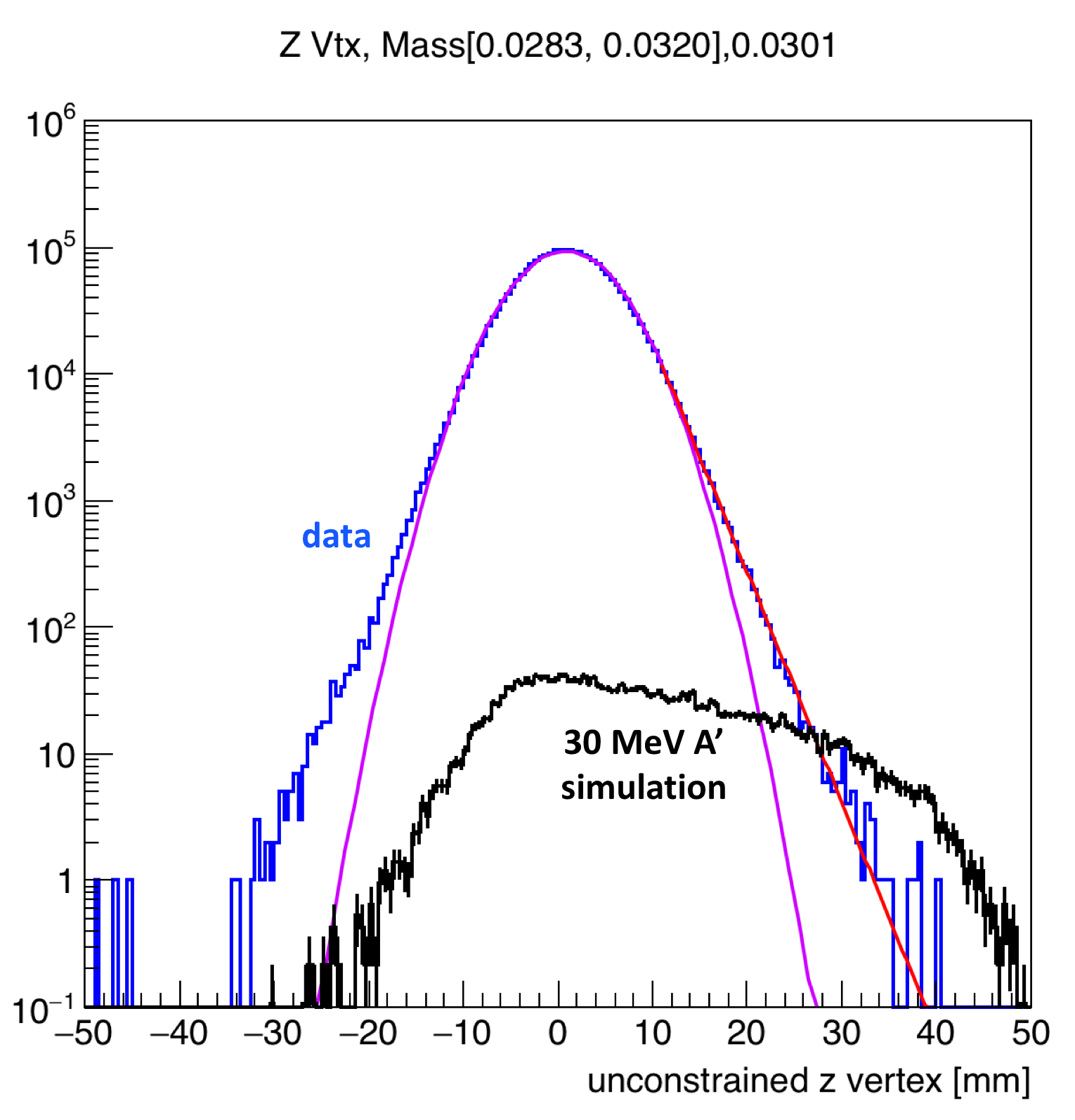}
            \includegraphics[width=.50\linewidth]{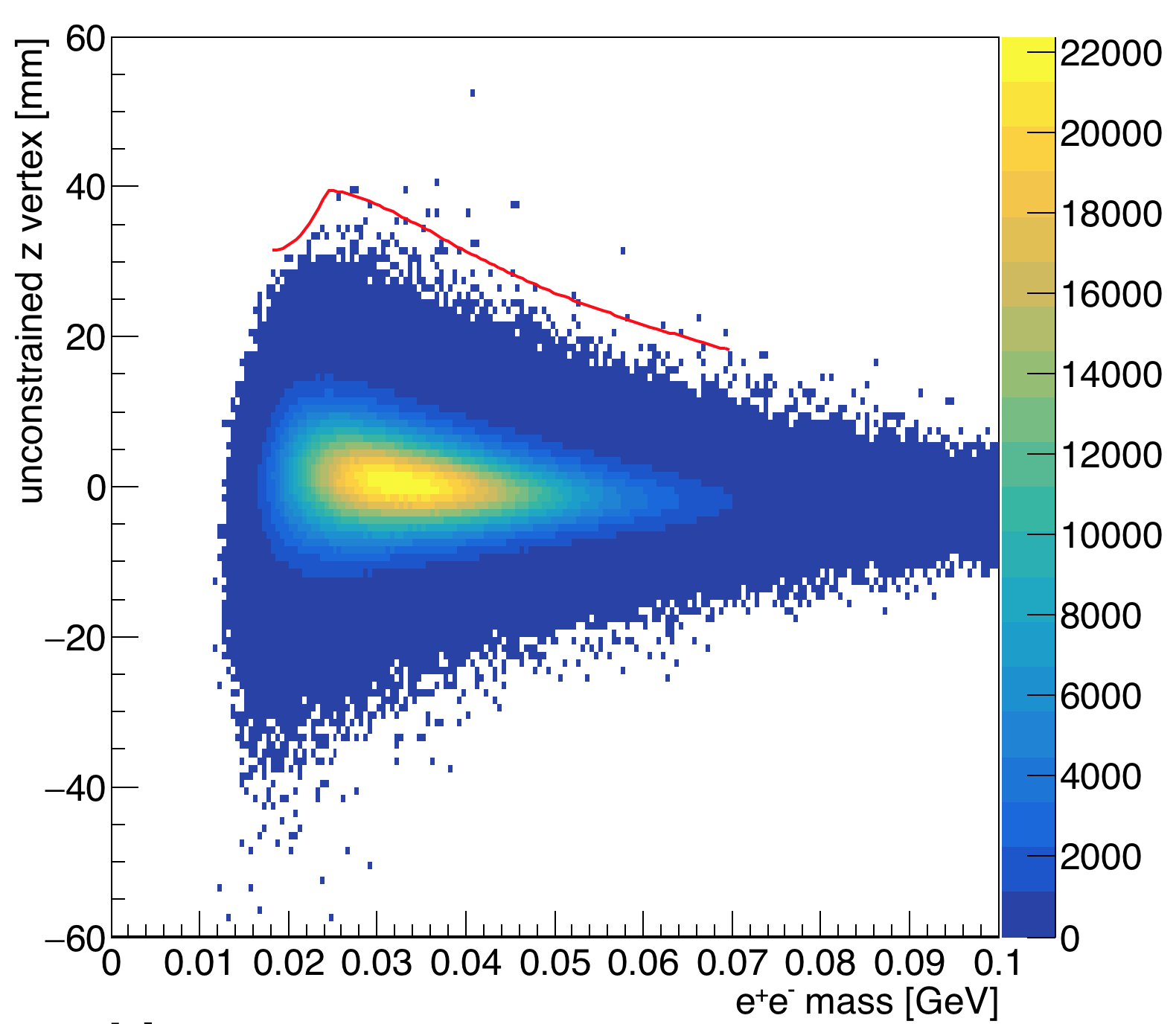}
            \caption{
                Left: The vertex distribution for a mass slice from data (blue) with all cuts applied and with Gaussian core fit (magenta) and downstream exponential tail fit (red) overlaid with a simulated $A^{\prime}$ Monte Carlo at the same mass (black). The tail distribution of the $A^{\prime}$ falls off more slowly compared to the distribution of events and scattering tails from the target but falls off rapidly beyond 40~mm downstream due to acceptance effects of requiring hits in Layer 1. For this particular mass slice, the $z_{cut}$ was found to be at 37.6~mm.
                Right: The reconstructed $z$ vertex is shown versus the reconstructed mass of the $e^+e^-$ pair for all selected events that in the 2015 data set. The $z_{cut}$ is obtained by slicing this distribution by mass hypothesis (window size of $\pm1.9\sigma_{m}$) and fitting each $z$ vertex distribution with a Gaussian with an exponential tail. The $z_{cut}$ shown in red is the point beyond which 0.5 background events is expected based on the fit to the tail.}
            \label{fig:zvtx}
        \end{figure}

In addition, both tracks were required to have layer 1 and layer 2 SVT hits. The layer 1 requirement removes tracks that have degraded vertex resolution and particles that may have scattered in the dead region of the layer 1 silicon (which naturally reconstruct a false vertex downstream of the target). The layer 1 requirement also has a significant effect on the reconstruction efficiency for long-lived heavy photons as shown in the $A^{\prime}$ distribution in Figure~\ref{fig:zvtx}.

        The heavy photon production rate (cross section), for both prompt and displaced vertices, is proportional to the radiative trident cross section. In looking for heavy photons with displaced vertices, the vertex distribution is examined in bins of the reconstructed invariant mass of the $e^+e^-$ pair over the full acceptance as shown in Figure~\ref{fig:zvtx}, and events of interest are identified as originating far beyond the tails of the prompt trident backgrounds. 

When searching for a displaced heavy photon, a downstream region having little background must be selected. Therefore, a downstream $z$ vertex position beyond which there should be fewer than 0.5~background events per mass bin, called a $z_{cut}$, was chosen. The $z_{cut}$ varies as a function of mass as shown in red in Figure~\ref{fig:zvtx}. Based on Poisson statistics alone, the 90$\%$ confidence limit for zero background requires us to have an expected number of $A^{\prime}$ events greater than 2.3.\\

An upper limit on the heavy photon production at a given $m_{A'}$ and $\epsilon^2$ is the maximum rate at which heavy photons could be produced, and still be consistent with the data. The confidence level used for this analysis is 90\%: in other words, if a heavy photon signal does exist at a given rate, the limit set by this analysis will (incorrectly) exclude that signal rate only 10\% of the time.

The method chosen for setting limits is the ``optimum interval'' method by Yellin \cite{Yellin:2002xd}.
This method was developed for dark matter direct detection experiments, and is intended for experiments where the signal shape is known, but the backgrounds are not fully understood and there is the possibility of an unexpected background. A particular strength of the method is that it minimizes the influence of a background that is concentrated in one part of the data distribution. 

\begin{figure}[th]
            \centering
            \includegraphics[width=1.0\linewidth]{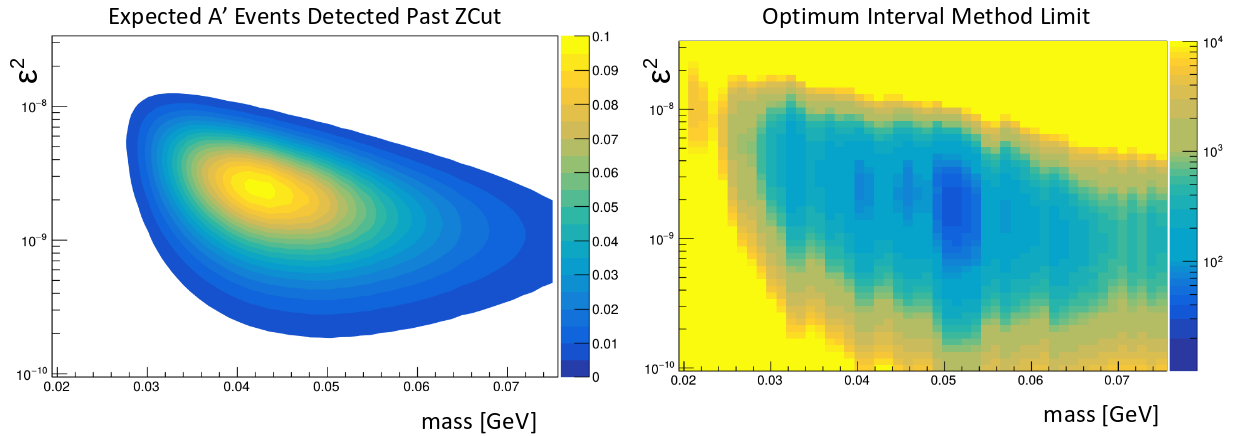}
            \caption{
                Left: The number of detectable $A^{\prime}$ events after applying all the analysis cuts and integrating the signal shape in Figure~\ref{fig:zvtx} over $z>z_{cut}$. The number of detectable $A^{\prime}$s is found to be a maximum of $0.097$ events where $A^{\prime}$ production is maximal at a mass of 43.6 MeV and $\epsilon^2$ coupling of $2.4\times 10^{-9}$. Right: 90\% confidence level upper limit on $\mu/\mu_{exp}$, the ratio of the true production rate to the expected production rate for a heavy photon. A value of 1 would mean exclusion; the lowest contour on this plot is 35.7 (which can be interpreted as just beginning to exclude an $\aprime$-like model with 35.7 times the cross section) at a mass of 51.4~MeV and coupling of $1.7\times 10^{-9}$. The vertical ridges in this plot correspond to the locations of events in mass space.}
            \label{fig:results}
        \end{figure}

The optimum interval method sets a one-sided upper limit (with confidence level $C$) on the number of signal events $\mu$ in a one-dimensional data set, where the shape of the signal distribution is known.
For HPS, the data set is the distribution of vertex $z$ locations, after applying the mass (in mass bins proportional to the mass resolution) and $z_{cut}$ cuts; the signal shape is the distribution reported in Figure ~\ref{fig:zvtx} for the $m_{A'}$ and $\epsilon^2$ being tested. The maximum number of detectable $A^{\prime}$s and the corresponding limit as determined by the optimum interval method is shown in Fig.~\ref{fig:results}.

By definition of $z_{cut}$, about 0.5 background events past $z_{cut}$ per mass slice (about 4 total) are expected; however, Figure~\ref{fig:zvtx} show many more such events past $z_{cut}$. This indicates that our background model is not complete in describing the background events. Detailed Monte Carlo studies reveal a similar deviation from the background model with some events due to mis-hits in layer 1 (which the isolation cut fails to eliminate) and large scatters due to a single Coulomb interaction (as opposed to multiple scattering) in the layer 1 silicon. Further work will need to be done to mitigate these backgrounds and better model them, which will be incorporated in future analysis.

\section{Conclusion}\label{sec:conclusion}

        Both a resonance search and a displaced vertex search for a heavy photon with a mass between 19 and 81 MeV which decays to an $\epem$ pair were performed.  A search for a resonance in
        the $\epem$ invariant mass spectrum yielded no significant excess and 
        established upper limits on the square of the coupling at the level of
        $6\times10^{-6}$, confirming results of earlier searches. A search for displaced vertices has been established, although no heavy photon parameter space could be excluded. While not
        covering new territory in both searches in this short engineering run, these searches did
        establish that HPS operates as designed and will, with future running,
        extend coverage for $\epsilon^2$ below the level of 10$^{-6}$ in the resonance search and provide coverage of unexplored parameter space at smaller values of the coupling from a search for displaced vertices.

\section{Acknowledgments}\label{sec:acknowledgments}

        The authors are grateful for the outstanding efforts of the Jefferson 
        Laboratory Accelerator Division and the Hall B engineering group in 
        support of HPS. The research reported here is supported by the U.S.
        Department of Energy Office of Science, Office of Nuclear Physics, 
        Office of High Energy Physics, the French Centre National de la 
        Recherche Scientifique, 
        United Kingdom's Science and Technology Facilities Council (STFC),
        the Sesame project HPS@JLab funded by the French region Ile-de-France 
        and the Italian Istituto Nazionale di Fisica Nucleare. Jefferson Science
        Associates, LLC, operates the Thomas Jefferson National Accelerator
        Facility for the United States Department of Energy under Contract
        No. DE-AC05-060R23177.

\end{document}